\documentclass[twocolumn,aps,superscriptaddress,prl]{revtex4}

 \usepackage{caption}
 \usepackage[labelfont=bf,labelsep=quad,justification=justified]{caption}
 \usepackage{upgreek}
 \usepackage[normalem]{ulem}
\usepackage[dvipsnames]{xcolor}			
\usepackage{graphicx}

\usepackage{times}
\usepackage{amsmath}
\usepackage{amssymb}
\usepackage{units}
\usepackage{stmaryrd} 

\DeclareGraphicsExtensions{.pdf,.png,.eps,.jpg}

\begin{document}

\title{Light-induced renormalization of the band structure of chiral tellurium}

\author{G. Gatti}  \email[E-mail address: ]{gianmarco.gatti@unige.ch}
\affiliation{Department of Quantum Matter Physics, Universit\'e de Gen\`eve, CH-1211 Gen\`eve, Switzerland} 

\author{N. Tancogne-Dejean} 
\affiliation{Max Planck Institute for the Structure and Dynamics of Matter and Center for Free-Electron Laser Science, Luruper Chaussee 149, 22761 Hamburg, Germany}

\author{H. H\"ubener}
\affiliation{Max Planck Institute for the Structure and Dynamics of Matter and Center for Free-Electron Laser Science, Luruper Chaussee 149, 22761 Hamburg, Germany}

\author{U. De Giovannini}
\affiliation{Max Planck Institute for the Structure and Dynamics of Matter and Center for Free-Electron Laser Science, Luruper Chaussee 149, 22761 Hamburg, Germany}
\affiliation{Dipartimento di Fisica e Chimica, Universit\'a degli Studi di Palermo, Via Archirafi 36, 90123 Palermo, Italy}

\author{J. Dai} 
\affiliation{Institute of Physics, \'Ecole Polytechnique F\'ed\'erale de Lausanne (EPFL), CH-1015 Lausanne, Switzerland} 
\affiliation{Lausanne Centre for Ultrafast Science (LACUS), \'Ecole Polytechnique F\'ed\'erale de Lausanne (EPFL), CH-1015 Lausanne, Switzerland} 
\affiliation{ALBA Synchrotron Light Source, Cerdanyola del Vall\`es, 08290 Barcelona, 
Spain}

\author{S. Polishchuk} 
\affiliation{Laboratory of Ultrafast Spectroscopy, ISIC, \'Ecole Polytechnique F\'ed\'erale de Lausanne (EPFL), CH-1015 Lausanne, Switzerland} 
\affiliation{Lausanne Centre for Ultrafast Science (LACUS), \'Ecole Polytechnique F\'ed\'erale de Lausanne (EPFL), CH-1015 Lausanne, Switzerland} 

\author{Ph. Bugnon}
\affiliation{Institute of Physics, \'Ecole Polytechnique F\'ed\'erale de Lausanne (EPFL), CH-1015 Lausanne, Switzerland} 

\author{F. Frassetto} 
\affiliation{National Research Council of Italy - Institute of Photonics and Nanotechnologies (CNR-IFN), via Trasea 7, 35131 Padova, Italy}

\author{L. Poletto} 
\affiliation{National Research Council of Italy - Institute of Photonics and Nanotechnologies (CNR-IFN), via Trasea 7, 35131 Padova, Italy}

\author{M. Chergui} 
\affiliation{Laboratory of Ultrafast Spectroscopy, ISIC, \'Ecole Polytechnique F\'ed\'erale de Lausanne (EPFL), CH-1015 Lausanne, Switzerland} 
\affiliation{Lausanne Centre for Ultrafast Science (LACUS), \'Ecole Polytechnique F\'ed\'erale de Lausanne (EPFL), CH-1015 Lausanne, Switzerland} 

\author{M. Grioni} 
\affiliation{Institute of Physics, \'Ecole Polytechnique F\'ed\'erale de Lausanne (EPFL), CH-1015 Lausanne, Switzerland} 
\affiliation{Lausanne Centre for Ultrafast Science (LACUS), \'Ecole Polytechnique F\'ed\'erale de Lausanne (EPFL), CH-1015 Lausanne, Switzerland} 

\author{A. Rubio} 
\affiliation{Max Planck Institute for the Structure and Dynamics of Matter and Center for Free-Electron Laser Science, Luruper Chaussee 149, 22761 Hamburg, Germany}
\affiliation{Nano-Bio Spectroscopy Group, Departamento de Fisica de Materiales, Universidad del Pa\'is Vasco, 20018 San Sebastian, Spain}
\affiliation{Center for Computational Quantum Physics (CCQ), The Flatiron Institute, 162 Fifth avenue, New York, New York 10010, USA}

\author{M. Puppin} \email[E-mail address: ]{michele.puppin@epfl.ch}
\affiliation{Laboratory of Ultrafast Spectroscopy, ISIC, \'Ecole Polytechnique F\'ed\'erale de Lausanne (EPFL), CH-1015 Lausanne, Switzerland} 
\affiliation{Lausanne Centre for Ultrafast Science (LACUS), \'Ecole Polytechnique F\'ed\'erale de Lausanne (EPFL), CH-1015 Lausanne, Switzerland} 

\author{A. Crepaldi} \email[E-mail address: ]{alberto.crepaldi@polimi.it}
\affiliation{Dipartimento di Fisica, Politecnico di Milano, Piazza Leonardo da Vinci 32, Milan 20133, Italy} 

\date{\today}


\begin{abstract}

Chirality in tellurium derives from a Peierls distortion driven by strong electron-phonon coupling, making this material a unique candidate for observing a light-induced topological phase transition. By using time- and angle-resolved photoelectron spectroscopy (trARPES), we reveal that upon near-infrared photoexcitation the Peierls gap is modulated by displacively excited coherent phonons with $\mathrm{A_{1g}}$ symmetry as well as chiral-symmetry-breaking $\mathrm{E'_{LO}}$ modes. 
By comparison with state-of-the-art TDDFT+U calculations, we reveal the microscopic origin of the in-phase oscillations of band edges, due to phonon-induced modulation of the effective Hubbard $U$ term.

\end{abstract}

\maketitle


Several strategies have been successfully put in action to optically control the electronic properties of solids, and induce new phases of matter \cite{Fausti_Science, Ichikawa_NM_11, Mihailovic_Science_14}, by leveraging the lattice \cite{Forst_NP_11} and the electronic degrees of freedom \cite{Tancogne_PRL_18, beaulieu2021ultrafast, topp2018all, Gatti_PRL_20, Marsi_PRR_20, Huber_SciAdv_24}. Bulk trigonal Te offers an ideal playground to assess the weight of those two ingredients, as the material is characterized by strong electron-phonon coupling bearing a Peierls distortion \cite{Tangney_PRL_99}. The opening of a small semiconducting band-gap ($\Delta E\sim$\,0.33\,eV \cite{Caldwell_PRB_59, Tutihasi_PR_69}) counterbalances the energy paid in the lattice distortion that stabilizes the chiral crystal structure. This delicate energy balance is sensitive to external perturbations and thereby explains the complex pressure-dependent phase diagram, which exhibits several changes in symmetry \cite{Bridg_38, Bridg_40, Bridg_45, Jamieson_JCP_65, Partha_PRB_88}. 


In this Letter we investigate the effects of an optical excitation on the band structure of bulk Te. By exploiting time- and angle-resolved photoelectron spectroscopy (trARPES) we report a band-gap renormalization (BGR) as large as 80\,meV. The ultrafast BGR is accompanied by a dynamical modulation of the binding energies of the valence (VB) and conduction bands (CB) that oscillate with maximum amplitude of $\sim 15$\,meV. According to our analysis in the time and frequency domains, both the $A_{1g}$ mode at 3.46\,THz and the $E'_{LO}$ mode at 2.97\,THz contribute to renormalize the band structure. Ab initio calculations reproduce the displacive excitation of the $A_{1g}$ mode and the in-phase oscillation of the band edges. 
Experimental observation of the $E'_{LO}$ coherent phonon in the modulation of the band structure indicates the possibility to excite symmetry breaking modes to modify the chiral structure of Te thus manipulating the Weyl points, as proposed for other Weyl semimetals \cite{Sie_Nature_19, Zhang_PRX_19, Hein_NC_20}.


Single crystals of Te were synthesized by the temperature gradient method \cite{Gatti_Te_PRL_20}. Post-cleavage in ultra-high vacuum leaves large atomically clean terraces ideal for trARPES experiment. The band structure was previously characterized by synchrotron-based ARPES and spin-resolved ARPES, as reported elsewhere \cite{Gatti_Te_PRL_20}. Time-resolved experiments were carried out at the solid-state end station of the Harmonium facility \cite{Ojeda_sd_16, Crepaldi_Chimia_17}, part of LACUS at EPFL, with energy, angular and temporal resolutions equal to 180\,meV, 0.3$^{\circ}$ and 60\,fs respectively. During measurements, the sample was at room temperature, and photoelectrons were emitted by s-polarized probe light with 17.6 eV photon energy, corresponding to the 11th harmonic generated in argon. The optical excitation was s-polarized and centered at 1.6 eV, with an absorbed fluence of $1.2\pm0.1$\,mJ/cm$^2$. 

We performed ab initio simulations using a first principles TDDFT+U framework \cite{Tancogne_PRB_2017} coupled to ion dynamics at the level of the Ehrenfest dynamics. At equilibrium, the effective U is calculated to be $U_{\rm eff} = 3.4$\,eV for the $5p$ orbitals of Te, yielding a band-gap of 0.355\,eV. More details on the ab initio simulations are given in Supplementary Note 5 \cite{SM}.



\begin{figure}[tt!]
  \centering   \includegraphics[width = 0.45 \textwidth]{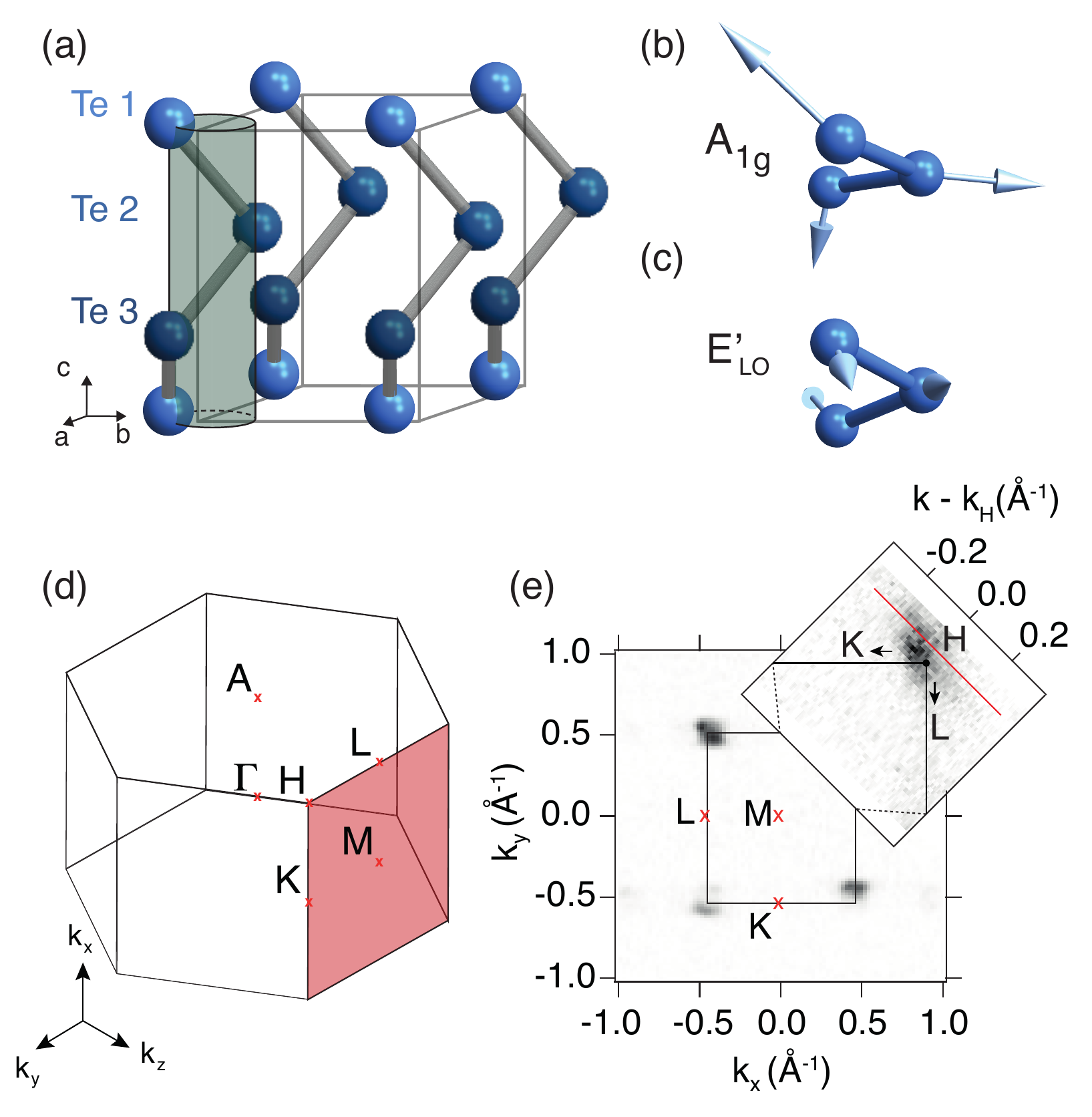}
  \captionsetup{justification=raggedright,singlelinecheck=false}
  \caption[didascalia]{ (a) Unit cell of Te with space group $P_{3_121}$ and lattice parameters \textit{a} = \textit{b} = 4.44\,$\mathrm{\AA}$, \textit{c} = 5.91\,$\mathrm{\AA}$ \cite{Bradley_24}. Different colors highlight the three atoms forming the helix along the $c$ axis. (b),(c) Eigenmodes of the $E'_{LO}$ and the $A_{1g}$ phonons, respectively. (d) 3D Brillouin zone along with the high symmetry points. (e) Fermi surface measured at 120\,eV and corresponding to the MLHK plane highlighted by a red rectangle in panel (d). The magnified view shows the region measured for a photon energy equal to 17.6 eV.}
  \label{fig:ARPES}
\end{figure}


Figure\,1\,(a) shows the crystal structure of the right-handed enantiomer of Te, that belongs to space group 152 ($P_{3_121}$) and whose unit cell contains four helices running along the $c$ crystallographic axis. Colors highlight the inequivalent Te atoms forming each chain, whose positions are related by the $C_{3_1}$ screw symmetry. The chiral symmetry is responsible for unparalleled magnetoelectric \cite{Tsirkin_PRB_18, Sahin_PRB_18, Rikken_PRB_19, Furukawa_NatCom_17, Moore_NatCom_16} and magneto-optical responses \cite{Asnin_JETP_78}, the formation of bound state realizing the Su–Schrieffer–Heeger model \cite{Nakayama_Nat_24} and exotic Weyl fermions \cite{Manes_PRB_12, Agapito_PRL_13, Hirayama_PRL_15, Tsirkin_PRB_17, Chang_NatMat_18, Chan_PRM_19} as well as their radial spin texture \cite{Sakano_PRL_20, Gatti_Te_PRL_20}.

Strong electron-phonon coupling not only underlies the intrinsic chirality of Te, but also governs its dynamical response to an ultrafast optical excitation. In fact, Te has been a paradigm for the study of coherent phonon excitation in opaque materials. In centrosymmetric non-polar crystals such as Bi and Sb, the generation mechanisms are the displacive excitation of coherent phonons (DECP) and the impulsive stimulated Raman scattering (ISRS)  \cite{Merlin_SSC_97, Nelson_ChemRev_94, Stevens_PRB_02}.  The broken inversion symmetry in Te also allows the efficient generation of longitudinal optical phonons by photo-Dember effect \cite{Dekorsy_EPL_93, Kuznetov_PRB_95, Kamaraju_PRB_10,  Huber_APL_15}. The optically induced macroscopic polarization is large enough to account for THz emission \cite{Dekorsy_PRL_95}, and time resolved optics revealed that in Te the amplitudes of the $A_{1g}$ and the $E'_{LO}$ phonons are comparable \cite{Micochko_JPCM_05}. Figure\,1\,(b) and (c)  show cartoons of two corresponding eigenmodes \cite{Inan_PR_68, Pine_PRB_71, Powell_75}. The $A_{1g}$  phonon is a breathing mode that increases the helix radius, indicated by a colored cylinder in Fig.\,1\,(a) \cite{Tangney_PRL_99, Johnson_PRL_09}. The $E'_{LO}$ mode breaks the $C_{3_1}$ screw symmetry [Fig.\,1\,(c)] \cite{Micochko_JPCM_05, Huber_APL_15}. Therefore coherent phonons excitation offers multiple approaches to alter the structure and symmetry of Te. 
 
The VB of Te is composed by two manifolds; the one at low-energy derives from the Te bonding orbitals, while the second, closer to the Fermi level ($\mathrm{E_F}$), is formed by lone pairs electrons. Finally, anti-bonding orbitals are responsible for the unoccupied CB \cite{Hirayama_PRL_15}. Previous ARPES experiments have carefully addressed the dispersion of the occupied bands over the three-dimensional Brillouin zone \cite{Nakayama_PRB_17, Gatti_Te_PRL_20}, which is displayed in Fig.\,1\,(d) along with the high-symmetry points. Cleavage exposes the $\mathrm{(10\bar{1}0)}$ surface and different high symmetry planes are accessible by varying the probe photon energy, and the value of the wave vector component along the surface normal, $k_z$, is retrieved for an inner potential V$_0 = 10$\,eV \cite{Gatti_Te_PRL_20}. In this study we focus on the MLHK plane, highlighted by a red rectangle in Fig.\,1\,(d). The corresponding Fermi surface is displayed in Fig.\,1\,(e) for a photon energy of 120 eV, while the inset shows the results obtained at 17.6 eV.  The Fermi surface consists in a tiny hole-like pocket centered at the H point in agreement with previous studies \cite{Sakano_PRL_20, Gatti_Te_PRL_20}.  A red line traces the cut where we follow the dynamical change in the band structure upon optical excitation. 


\begin{figure*}[tt!]
  \centering   \includegraphics[width = 0.9 \textwidth]{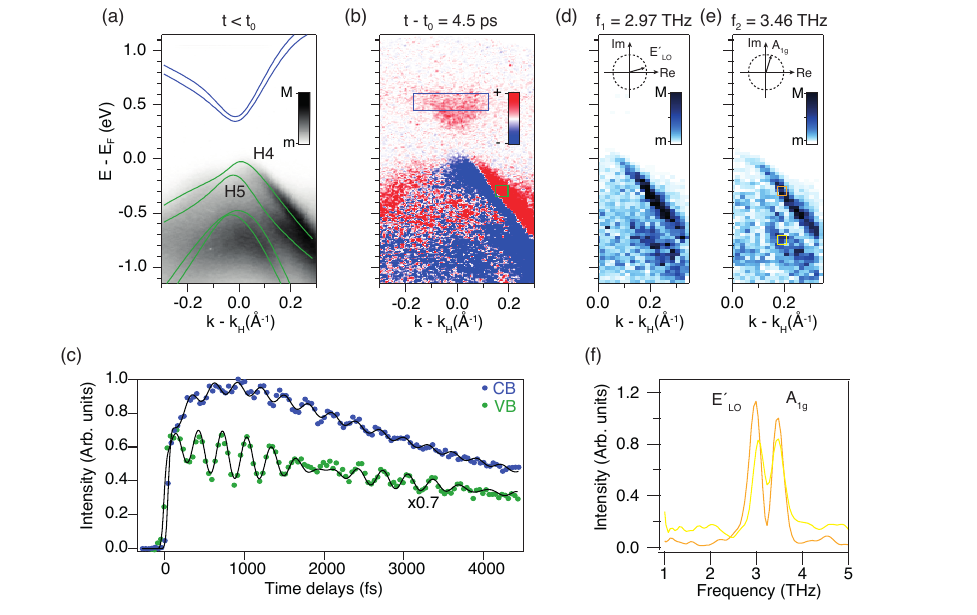}
  \captionsetup{justification=raggedright,singlelinecheck=false}
  \caption[didascalia]{ (a) Reference band structure measured before optical excitation ($t<t_0$) along the direction passing near H, indicated by the red line in Fig.1 (e).  Green and blue curves correspond to the equilibrium TDDFT+U VB and CB, respectively, obtained with an effective calculated $U_{\rm eff}$ of 3.4\, eV. (b) Differential image obtained as difference between the ARPES data at $\mathrm{t-t_0 = 4.5}$\,ps and the reference. (c) Dynamical evolution of the intensity integrated in the rectangles at the edges of VB and CB, as shown in panel (b); signal in VB is multiplied by factor 0.7 to improve readability of the curves. (d),(e) Fourier amplitude images corresponding to the components at 2.97$\pm$\,0.05\,THz (d) and 3.46$\pm$\,0.06\,THz (e), insets show their relative phases in the complex plane. (f) Fourier spectra corresponding to the region in bands H4 and H5 indicated by colored rectangles in panel (e).}
  \label{fig:tr_ARPES}
\end{figure*}



Figure\,2\,(a) shows a reference ARPES image before the arrival of the pump, at $t < t_0$.  A suppression of spectral weight is apparent for $k \leqslant k_\mathrm{H}$ due to matrix element effects. The two highest occupied bands are labeled H4 and H5 according to their irreducible representation, green and blue lines trace the calculated DFT+U band dispersion that is in good agreement with the experiment. Figure\,2\,(b) reveals the prominent effects of the optical excitation. The image is obtained as difference between the ARPES data at $\mathrm{t-t_0 = 4.5}$\,ps and the reference signal at $t < t_0$. A complete movie of electron dynamics in the range between $-0.3$\,ps and $4.5$\,ps can be found in the Supplementary Material \cite{SM}. Red and blue colors encode increase and decrease of the photoemission signal, respectively. Upon optical excitation we observe a transfer of electrons to CB. They subsequently scatter to the band minimum through cascade processes over a time scale of $0.43\pm0.05$\,ps. The blue-red signal in correspondence of the H4 band is a consequence of the energy shift of VB, the signature of the BGR. 

Figure\,2\,(c) shows the temporal evolution of the signal at the edges of VB (green) and CB (blue), integrated in the regions indicated by the rectangles in Fig.\,2\,(b), along with the best fits (black lines). Oscillations are clearly visible, overlaid to an exponential decay with characteristic time of $\tau_{VB} = 6.6\pm0.8$\,ps and $\tau_{CB}= 4.4\pm0.2$\,ps, for VB and CB respectively. The signal in VB exhibits beatings, which point towards the simultaneous presence of two modes with close values of the periods. In the analysis, those modes are accounted by two sine functions with characteristic frequencies of 3.46\,$\pm$\,0.06\,THz and 2.97\,$\pm$\,0.05\,THz, slightly smaller than the literature values for the $A_{1g}$ (3.6\,THz) and the longitudinal optical $E'_{LO}$ modes (3.09\,THz) \cite{Dekorsy_PRL_95, Micochko_JPCM_05, Huber_APL_15}, thus indicating a softening of the phonons upon optical excitation (see Supplementary Note 1 for a detailed description of the fit \cite{SM}). The two modes exhibit different initial phases, $1.29\pm0.08$\,rad for the $A_{1g}$ and $0.28\pm0.18$\,rad for the  $E'_{LO}$. The former reflects the DECP excitation mechanism: the charges accumulated in the CB weaken the bonds among the atoms of the helix, whose radius is free to expand. As a results, the oscillation of the ions around the new quasi-static positions is well accounted for by a cosine function \cite{Merlin_SSC_97, Tangney_PRB_02, Kudryashov_PRB_07}. The phase of the $E'_{LO}$ mode reflects the sinusoidal oscillations commonly observed for the longitudinal optical phonons excited by the photo-Dember effects \cite{Cho_PRL_90}. This mechanism is possible because the mode is also infrared active, and it can be stimulated by an optical perturbation with an electric field component orthogonal to the $c$ axis, as in our experimental geometry. 

In order to evaluate how the two modes couple to the band structure, we perform a mode selective Fourier amplitude analysis of the measured tr-ARPES data \cite{DeGiovannini_PRL_20, Hein_NC_20} (see Supplementary Note 2 for a detailed description of the procedure \cite{SM}). Figure\,2\,(d) and (e) shows maps corresponding to 2.97\,THz and 3.46\,THz, respectively. The Fourier analysis confirms the difference in-phase between the two modes, as shown by phasors in the insets. Figure\,2\,(f) shows the Fourier amplitude spectra for the regions corresponding to H4 and H5, enclosed by colored rectangles in Fig.\,2\,(e), and we conclude that the amplitude of the modes is similar over the entire VB dispersion.


  
\begin{figure*}[tt!]
  \centering   \includegraphics[width = 0.9 \textwidth]{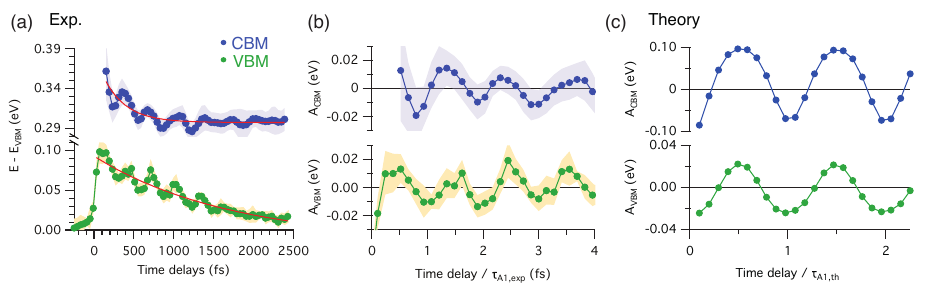}
  \captionsetup{justification=raggedright,singlelinecheck=false}
  \caption[didascalia]{(a) Energy positions of the conduction and valence band edges as a function of time delay, as extracted from tr-ARPES experiments. (b) Relative energy positions of the conduction and valence band obtained subtracting an exponential decay (red curve in panel a) from the traces of panel (a). The time axis is renormalized to the experimental value of the period of the $A_{1g}$ phonon mode $\tau_{A1,exp}$. Shaded areas indicate 3 standard deviations error bars}. (c)  Energy positions of the conduction and valence band edges as a function of time delay, as obtained from theory. The time axis is renormalized to the theoretical value of the period of the A$_{1g}$ phonon mode $\tau_{A1,th}$. 
  \label{fig:tr_ARPES}
\end{figure*}


We now assess quantitatively the light-induced BGR, by extracting energy distribution curves (EDCs) integrated over $\Delta k = \pm 0.1$\,\AA $^{-1}$ around the H point. For each time delay we identify the leading edges of VB and CB (see Supplementary Note 3 for a detailed description of the procedure \cite{SM}), and Figure\,3\,(a) shows the dynamics of their energy position, referenced to the equilibrium VB maximum (VBM). Within the probed time scale, the VB leading edge is a proxy of the VBM, which, upon optical excitation, shifts upward by $\sim 80$\,meV before relaxing back to its original position in 2.5 ps. This result is corroborated by a complementary analysis of the entire VB dispersion via momentum distribution curves, MDCs (see Supplementary Note 4 for a detailed description of the procedure \cite{SM}). The CB leading edge is representative of the CB minimum (CBM) once the electronic thermalization has occurred, which is after few hundreds of fs after time overlap. Similarly to the valence band, the dynamics of CBM is well approximated by an exponential decay, but with a faster decay constant. The difference between the two relaxation curves defines dynamically the size of the band-gap. At $t-t_0 =$\,0.3\,ps the gap is found to be $\sim 250$\,meV, corresponding to a BGR of $\sim 80$\,meV. This value is comparable with the VBM shift and suggests that the optical excitation does not induce a symmetric closure of the band-gap. 

Figure 3\,(b) displays the relative positions of the VBM and CBM with respect to their absolute energy displacement from equilibrium, described by an exponential decay (red curve in Fig.\,3\,(a)). The time-dependent BGR is modulated by periodic oscillations of $\sim 15$ meV amplitude in the binding energies of the two states. The two oscillations share the same phase, thus indicating that the periodic motion of the ions around the new quasi-equilibrium values induces a shift of VBM and CBM in the same direction. 
In order to analyze the experimental results, we performed TDDFT+U calculations with Ehrenfest nuclear dynamics, for an excitation fluence corresponding to the experimental one of $1.2\pm0.1$\,mJ/cm$^2$. 
From the knowledge of the atomic displacements for each time delay $\tau$, and the corresponding U($\tau$), we further performed a set of static band structure calculations at the level of DFT+U($\tau$) to extract the change in the band-gap, shown in Fig.\,3\,(c).
By comparing the U($\tau$) with the self-consistent U of these static calculations we find that the band-gap oscillations are exclusively results of the adiabatic modulations of U due to the ionic motion.

Theory qualitatively reproduces the in-phase oscillations of VBM and CBM, with a magnitude of $\sim$\,20 and $\sim$\,100 meV respectively, associated to the excitation of the  $A_{1g}$  mode with calculated frequency 3.7\,THz. This explains the microscopic origin of the in-phase oscillations of the band-gap, originating from the phonon-induced modulation of the Hubbard U term.
This calculation, that only accounts for structural contributions to the band-gap dynamics, does not capture the BGR of 80 meV that is experimentally observed immediately after optical excitation.

From the lattice point of view, it is well known that a variation of the helix radius from the equilibrium value of 0.26 \textit{a} (in lattice unit) to 1/3 \textit{a} induces a transition towards the rhombohedral $R_{ \overline{3}m}$ space group  \cite{Tangney_PRL_99, Kudryashov_PRB_07}. This change in symmetry is predicted by theory to be accompanied by a closure of the band-gap, and a possible topological phase transition as recently proposed on the basis of TDDFT and time-resolved second harmonic generation for a critical fluence of $\sim$\,2\,mJ/cm$^2$ \cite{Ning_PRB_22}. Based on our results we conclude that slightly below the critical fluence for above band-gap excitation, the BGR due to the ionic motion alone seems to be insufficient to close the band-gap. This suggests that to achieve symmetry and topological phase transitions it will be necessary to implement strategies to selectively couple the optical excitation with lattice distortions, as for example by tuning the pump energy in the mid-infrared and terahertz range to resonantly excite specific phonon modes.


Finally, we point out that the full closure of the band-gap is not observed in our trARPES data that are acquired with pump fluences below the threshold set by the emergence of pump induced artifacts. Nonetheless, other important topological changes can be controlled by the excitation of the $E'_{LO}$ phonon mode. In Te, in fact, several Weyl points are pinned at the H point by the $C_{3_1}$ axis \cite{Nakayama_PRB_17, Gatti_Te_PRL_20}, and once this symmetry is broken by the coherent phonon, the band degeneracy is moved away from the high symmetry point. This effect offers, in our opinion, an alternative explanation to the results of a recent time-resolved optical study of the transient band-gap of Te nanosheets \cite{Jnawali_NC_20}. The authors report that upon optical excitation, new optical transitions around the band edges become visible due to the splitting of the Weyl point at CBM. The effect is interpreted in terms of shear strain induced in the film by light through the inverse piezoelectric effect \cite{Jnawali_NC_20}. But considering that the excitation energy and fluence are comparable to one used in our study, the excitation of $E'_{LO}$ mode might also explain the splitting of the Weyl point at H. In order to fully capture the optical excitation of the $E'_{LO}$ mode, and its consequences on the band structure of Te, TDDFT+U calculations should be combined with a crystal slab geometry to account for the photo-Dember effect. This is beyond the scope of our study, but we are confident that our experimental results will stimulate a future theoretical activity in this direction.


In conclusion, we observed that upon optical excitation the band structure of Te exhibits a band-gap renormalization as large as 80 meV, accompanied by in-phase oscillations of the valence and conduction band edges with amplitude of $\sim 15$ meV, and frequencies equal to 3.46\,$\pm$\,0.06\,THz and 2.97\,$\pm$\,0.05\,THz, associated to the excitation of symmetric $A_{1g}$ and symmetry breaking  $E'_{LO}$ modes, respectively. Ab initio TDDFT + U calculations reproduce oscillations in the band structure due to the $A_{1g}$ mode, but not the large initial value of the BGR. 

Finally, we propose that the excitation of a symmetry breaking $E'_{LO}$ might be responsible for the recent observation of optical transitions that become accessible when the Weyl points are moved away from the CBM at H. Strain and static lattice distortions have been already proposed as strategies to control the topological phase of Te by altering the crystal symmetry \cite{Agapito_PRL_13, Hirayama_PRL_15}. Our results enlarge these horizons by showing that intense optical excitation is a complementary route to control topological phase transitions in a chiral crystal by altering its symmetry, even before the full band-gap closure.



\section*{Acknowledgements}

Correspondence and requests for materials should be addressed to G.G. \,(gianmarco.gatti@unige.ch) M.P.\,(michele.puppin@epfl.ch) and A.C.\,(alberto.crepaldi@polimi.it).  This research used resources of the Advanced Light Source, which is a DOE Office of Science User Facility, under Contract No. DE-AC02-05CH11231
We acknowledge  financial support by the Swiss National Science Foundation (SNSF), via the NCCR:MUST and the contracts No. 206021-157773, and 407040-154056 (PNR 70). This work was supported by the ERC Advanced Grant H2020 ERCEA 695197 DYNAMOX. We acknowledge financial support from the European Research Council (ERC-2015-AdG-694097). The Flatiron Institute is a division of the Simons Foundation. This work was supported by the Cluster of Excellence Advanced Imaging of Matter (AIM), Grupos Consolidados (IT1249-19) and SFB925.





\end{document}